\documentclass[twoside,12pt]{article}
\usepackage{epsfig,,amsmath,amssymb}

\def\Journal#1#2#3#4{{#1} {#2} (#4) #3 }

\newcommand{\be}{\begin{equation}}
\newcommand{\ee}{\end{equation}}
\newcommand{\bea}{\begin{eqnarray}}
\newcommand{\eea}{\end{eqnarray}}

\newcommand{\sfrac}[2]{\mbox{\footnotesize $\displaystyle \frac{#1}{#2}$}}

\begin{document}

\title{Hadron Properties and Dyson-Schwinger Equations}
\author{C.\,D.\ Roberts\\ \\ Physics Division, 
Argonne National Laboratory,
Argonne IL 60439, USA}
\maketitle

\begin{abstract} 
An overview of the theory and phenomenology of hadrons and QCD is provided from a Dyson-Schwinger equation viewpoint.  Following a discussion of the definition and realisation of light-quark confinement, the nonperturbative nature of the running mass in QCD and inferences from the gap equation relating to the radius of convergence for expansions of observables in the current-quark mass are described.  Some exact results for pseudoscalar mesons are also highlighted, with details relating to the $U_A(1)$ problem, and calculated masses of the lightest $J=0,1$ states are discussed.  Studies of nucleon properties are recapitulated upon and illustrated: through a comparison of the $\ln$-weighted ratios of Pauli and Dirac form factors for the neutron and proton; and a perspective on the contribution of quark orbital angular momentum to the spin of a nucleon at rest.  Comments on prospects for the future of the study of quarks in hadrons and nuclei round out the contribution. \end{abstract}

\section{Introduction}
In trying to elucidate the role of quarks in hadrons and nuclei one steps immediately into the domain of relativistic quantum field theory where within the key phenomena can only be understood via nonperturbative methods.  Two prime examples are: confinement, the empirical fact that quarks have not hitherto been detected in isolation; and dynamical chiral symmetry breaking, which is responsible, amongst many other things, for the large mass splitting between parity partners in the spectrum of light-quark hadrons, even though the relevant current-quark masses are small.  Neither of these phenomena is apparent in QCD's Lagrangian and yet they play a dominant role in determining the observable characteristics of real-world QCD.  The physics of hadrons is ruled by such \emph{emergent phenomena}.

\section{Confinement}
\label{Sect:Conf}
In connection with confinement it is worth emphasising at the outset that the potential between infinitely-heavy quarks measured in numerical simulations of quenched lattice-regularised QCD -- the so-called static potential -- is simply not relevant to the question of light-quark confinement.  In fact, it is quite likely a basic feature of QCD that a quantum mechanical potential between light-quarks is impossible to speak of because particle creation and annihilation effects are essentially nonperturbative.  

A perspective on confinement was laid out in Ref.\,\cite{Krein:1990sf}.  Expressed simply, confinement can be related to the analytic properties of QCD's Schwinger functions, which are often loosely called Euclidean-space Green functions.  For example, it can be read from the reconstruction theorem that the only Schwinger functions which can be associated with expectation values in the Hilbert space of observables; namely, the set of measurable expectation values, are those that satisfy the axiom of reflection positivity \cite{gj81}.  This is an extremely tight constraint.  It can be shown to require as a necessary condition that the Fourier transform of the momentum-space Schwinger function is a positive-definite function of its arguments.  However, that is not sufficient.  

In relation to $2$-point Schwinger functions, which are those connected with the propagators of elementary excitations in QCD, the axiom of reflection positivity is satisfied if, and only if, the Schwinger function possesses a K\"all\'en-Lehmann representation.  This statement is most easily illustrated for a scalar field, in which case it means that one can write the scalar-field's $2$-point function in the form\footnote{A Euclidean metric will be used throughout.  In concrete terms that means: for Dirac matrices, $\{\gamma_\mu,\gamma_\nu\} = 2\delta_{\mu\nu}$, $\gamma_\mu^\dagger = \gamma_\mu$; and $a \cdot b = \sum_{i=1}^4 a_i b_i$.  A timelike vector, $p_\mu$, has $p^2<0$.  Naturally, no theory consistent with causality can produce a Schwinger function with a pole at spacelike $p^2$.}
\begin{equation}
\label{KLrep}
{\cal S}(p^2) = \int_0^\infty\! d\varsigma\, \frac{\rho(\varsigma)}{p^2+\varsigma^2}\,, \; \rho(\varsigma)\geq 0 \; \forall \varsigma \geq 0. 
\end{equation}
The spectral density for a non-interacting scalar field of mass $m$ is $\rho(\varsigma) = \delta(\varsigma-m)$.  It is plain that no function ${\cal S}(p^2)$ which falls-off faster than $1/p^2$ at large spacelike momenta can be expressed in this way.  

An appreciation of the importance of the axiom associated with reflection positivity has led to the formulation of a \emph{confinement test} \cite{Hawes:1993ef}.  With a momentum-space Schwinger function, ${\cal S}(p)$, in hand, the first step is to calculate
\begin{equation}
\label{Deltatau}
\Delta(\tau) = \int d^3 x \int\frac{d^4 p}{(2\pi)^4}\, {\rm e}^{i p_4 \tau + i \vec{p}\cdot\vec{x}}\, {\cal S}(p),
\end{equation}
which gives the configuration-space Schwinger function in the rest frame.\footnote{The arguments and conclusions that follow can readily be adapted to the case of models or theories without a mass gap; i.e., that possess a nonperturbatively massless excitation.}  One then examines the properties of $\Delta(\tau)$.  

If an asymptotic state of mass $m$ is associated with this Schwinger function, 
then 
\begin{equation}
\Delta(\tau) \stackrel{\tau\to \infty}{=} \frac{1}{2 m} \, {\rm e}^{- m \tau};
\end{equation}
i.e., the Schwinger function is positive definite and the mass of the asymptotic, propagating state is given by\footnote{This picture is readily generalised to the case of a Schwinger function that describes a channel with more than one asymptotic state; i.e., a ground state plus excitations \protect\cite{Bhagwat:2007rj}.}
\begin{equation}
- \lim_{\tau\to \infty} \, \frac{d}{d \tau}\ln \Delta(\tau) = m\,.
\label{massDelta}
\end{equation}
If, on the other hand, $\Delta(\tau)$ calculated from a particular Schwinger function is not positive definite, then the axiom of reflection positivity is violated and the associated elementary excitation does not appear in the Hilbert space of observables.  Thus the appearance of a least one zero in $\Delta(T)$ is a sufficient condition for confinement.  It is a very clear signal.

An exemplar is provided by 
\begin{equation}
\label{stingl}
{\cal S}(p) = \frac{p^2}{p^4+ 4 \mu^4}, 
\end{equation}
a function for which with any particular choice of $\epsilon> 0$ there exists a $p_\epsilon^2>0$ such that $\forall p^2 > p_\epsilon^2$ one has $|{\cal S}(p) - 1/p^2|<\epsilon$.  One calculates from Eq.\,(\ref{stingl}):
\begin{equation}
\Delta(\tau) = \frac{1}{4 \mu} \, {\rm e}^{- \mu \tau}\,\left[ \cos \mu \tau - \sin \mu \tau \right].
\end{equation}
This model was proposed in Ref.\,\cite{Stingl:1985hx} and argued therein to describe a confined gluon because it doesn't possess a pole on the real-$p^2$ axis.  The argument for confinement is supported by the fact that $\Delta(\tau)$ is not positive definite and, indeed, exhibits damped oscillations about zero.  Hence the excitation associated with this Schwinger function cannot be connected with an expectation value in the Hilbert space of observables.

The existence of a zero in $\Delta(\tau)$ and its disappearance have been advocated as a means by which deconfinement can be studied in QCD at nonzero temperature and density \cite{Bender:1996bm,Bender:1997jf}.  Deconfinement and chiral symmetry restoration are coincident in all self-consistent studies of concrete models of QCD that exhibit both phenomena in vacuum.  This result appears to follow from the crucial role played by the in-medium evolution of the dressed-quark self-energy in both transitions.  Further studies are needed to fully elucidate the nature and the probable simultaneity of these transitions in QCD.

It was observed and emphasised in Ref.\,\cite{Krein:1990sf} that a violation of reflection positivity is a sufficient but \emph{not} necessary condition for confinement.  This criterion is nevertheless a powerful discriminating tool, as can be illustrated through a concrete example, which also serves to introduce the Dyson-Schwinger equations (DSEs).  The DSEs provide a nonperturbative approach in the continuum for calculating and modelling the properties of Schwinger functions.  They have long been used to develop insight into gauge field theories at both weak and strong coupling \cite{Roberts:1994dr}.  

In connection with confinement and dynamical chiral symmetry breaking (DCSB) the DSE for the dressed-quark propagator is of fundamental importance.  In QCD this equation takes the form
\begin{equation}
S(p)^{-1} =  i\gamma\cdot p + m +  \int\frac{d^4 q}{(2\pi)^4} \, g^2 D_{\mu\nu}(p-q) \frac{\lambda^a}{2}\gamma_\mu S(q) \Gamma^a_\nu(q,p) , \label{gendseUR}
\end{equation}
where at this point it is sufficient to work with the unrenormalised equation.  In Eq.\,(\ref{gendseUR}), $D_{\mu\nu}$ and $\Gamma^a_\nu$ are the dressed-gluon 2-point function (propagator) and dressed-quark-gluon 3-point function (vertex); and $m$ is a current-quark bare mass.  Finally, the subject of the equation, $S(p)$, is the dressed-quark propagator in vacuum.  It can be written in a number of equivalent forms:
\begin{equation}
\label{SpAB}
S(p) = -i \gamma\cdot p \, \sigma_V(p^2) + \sigma_S(p^2) =  \frac{1}{i\gamma \cdot p \,A(p^2)+ B(p^2)} = \frac{Z(p^2)}{i\gamma \cdot p + M(p^2)},
\end{equation}
each of which involves just two scalar functions.  In the last representation they are chosen to be the wave function renormalisation, $Z(p^2)$, and the running mass function, $M(p^2)$.

Reference~\cite{Munczek:1983dx} introduced a simple \textit{Ansatz} for the kernel of Eq.\,(\ref{gendseUR}); namely,
\begin{equation}
\label{mnprop} g^2 D_{\mu\nu}(k) = (2\pi)^4\, {\cal G}\,\delta^4(k) \left[\delta_{\mu\nu} - \frac{k_\mu k_\nu}{k^2}\right],\; \Gamma^a_\nu(q,p) = \frac{\lambda^a}{2}\gamma_\nu\,,
\end{equation}
where ${\cal G}$ is a mass$^2$-scale.  The ladder quark-antiquark interaction represented by Eqs.\,(\ref{mnprop}) is a constant in configuration space.  Therefore the usual proof of the cluster decomposition property fails and hence this \textit{Ansatz} defines a confining model.  Furthermore, with $m=0$, Eqs.\,(\ref{gendseUR}) and (\ref{mnprop}) yield the following solution:
\begin{equation} 
\label{ABMNres}
\begin{array}{ll}
\begin{array}{ccc}
A(p^2) & =& \left\{ \begin{array}{ll} 
2\,;\; & p^2\leq {\cal G}\\ 
\frac{1}{2}\left( 1 + \sqrt{1+8{\cal G}/p^2} \right)\,;\; & p^2>{\cal G} \end{array} \right.,
\end{array}
\begin{array}{ccc}
B(p^2) &= & \left\{ \begin{array}{ll} 
2 \sqrt{{\cal G}-p^2}\,;\; & p^2\leq {\cal G}\\ 
0\,; & p^2>{\cal G} \end{array} \right. 
\end{array}.
\end{array}
\end{equation} 
When Eq.\,(\ref{Deltatau}) is evaluated with $\sigma_S(p^2)$ constructed from Eqs.\,(\ref{ABMNres}), one obtains
\begin{equation}
\label{DABMN}
\Delta(\tau) \propto \frac{J_1(\tau \surd {\cal G})}{\tau \surd {\cal G}}\,;
\end{equation}
i.e., a function with repeated zeros.  Thus the interaction's violation of the cluster decomposition property is expressed in a fermion propagator that fails to satisfy the axiom of reflection positivity.  Hence it describes a confined elementary excitation.

The story continues in a fascinating manner when one dresses the quark-gluon vertex.  Studies attempting to elucidate the nature of $\Gamma^a_\nu(q,p)$ are now underway using both continuum \cite{Hawes:1998cw,Bhagwat:2003vw,Bhagwat:2004hn,Bhagwat:2004kj,Matevosyan:2006bk} and lattice-QCD techniques \cite{Skullerud:2003qu}.  However, this is a relatively recent development.  It was and remains common to employ an \emph{Ansatz} for this 3-point Schwinger function that is constructed according to lessons learnt by studying Ward-Takahashi identities in gauge field theories \cite{Burden:1993gy,Dong:1994jr,Bashir:1994az}.  One often used form is based on the vertex introduced in Ref.\,\cite{Ball:1980ay}; viz.,\footnote{Improvements can be considered.  However, modifications that vanish at $\ell_1=\ell_2$ do not change the results which follow.  The influence of more significant changes; for example, introducing a dependence on functions other than those which appear in the quark propagator, must be explored on a case-by-case basis.  It is a simple truth that all we have at present are \textit{Ans\"atze} for the vertex and no modification will alter a simple, long established fact, soon to become apparent: the vertex can have a material impact on the expression of confinement and DCSB in QCD.} $\Gamma^a_\mu(\ell_1,\ell_2) = \frac{\lambda^a}{2} \,\Gamma_\mu(\ell_1,\ell_2)$, with 
\begin{eqnarray}
\label{bcvtx}
i\Gamma_\mu(\ell_1,\ell_2)  & =  &
i\Sigma_A(\ell_1^2,\ell_2^2)\,\gamma_\mu +
(\ell_1+\ell_2)_\mu \left[i\sfrac{1}{2}\gamma\cdot (\ell_1+\ell_2) \,
\Delta_A(\ell_1^2,\ell_2^2) + \Delta_B(\ell_1^2,\ell_2^2)\right] \!,\\
\Sigma_F(\ell_1^2,\ell_2^2)& = &\sfrac{1}{2}\,[F(\ell_1^2)+F(\ell_2^2)]\,,\;
\Delta_F(\ell_1^2,\ell_2^2) =
\frac{F(\ell_1^2)-F(\ell_2^2)}{\ell_1^2-\ell_2^2}\,,
\label{DeltaF}
\end{eqnarray}
where $F= A, B$; viz., the scalar functions in Eq.\,(\ref{SpAB}).  If one solves the gap equation, Eq.\,(\ref{gendseUR}), using this vertex model and $g^2 D_{\mu\nu}$ from Eq.\,(\ref{mnprop}) one finds in the chiral limit \cite{Burden:1991gd}
\begin{equation}
\label{VSMNres}
\sigma_V(p^2) = \frac{p^2 - 2 {\cal G} [1- \exp(-p^2/[2 {\cal G}])]}{p^4}\,,\;
\sigma_S(p^2) = \frac{1}{2\surd {\cal G}}\,\exp(-p^2/[2 {\cal G}])\,.
\end{equation}

It is apparent that dressing the vertex has markedly altered the form of the dressed-quark propagator; namely, $\sigma_{V,S}$, which each had a branch point with the bare vertex, are transformed by the dressed vertex into entire functions.  Moreover, when Eq.\,(\ref{Deltatau}) is evaluated with $\sigma_S(p^2)$ constructed from Eqs.\,(\ref{VSMNres}), one obtains
\begin{equation}
\label{DGT2}
\Delta(\tau) \propto \exp(-{\cal G}\tau^2/2)\,.
\end{equation}
In contrast to Eq.\,(\ref{DABMN}), this is positive definite.  However, it yields an effective mass via Eq.\,(\ref{massDelta}):
\begin{equation}
m = \lim_{\tau\to \infty} m(\tau) = \lim_{\tau\to \infty} {\cal G} \tau = \infty\,,
\end{equation}
and infinitely massive states \emph{cannot} propagate.  
%

These observations notwithstanding, confinement of the quark represented by the propagator in Eq.\,(\ref{VSMNres}) is still signalled by a violation of the reflection positivity axiom.  It is clear from the momentum space form of the propagator that the axiom is breached because both $\sigma_{V,S}$ fall faster than $1/p^2$ at large spacelike $p^2$.  However, it can also be seen in another way, which is indispensable if the momentum-space form of the Schwinger function is not known.  One can evaluate Eq.\,(\ref{Deltatau}) with Eq.\,(\ref{KLrep}) to obtain
\begin{equation}
\label{Widder}
\Delta(\tau) = \int_0^\infty \! d\varsigma \, \frac{\rho(\varsigma)}{2 \varsigma} \, {\rm e}^{-\varsigma \tau}\,.
\end{equation}
Reflection positivity is satisfied by a configuration-space Schwinger function if, and only if, it can be expressed in this form, with $\rho(\varsigma)$ as described in Eq.\,(\ref{KLrep}). Now the theorem of Ref.\,\cite{WTheorem} can be employed to show that the function in Eq.\,(\ref{DGT2}) cannot be represented in this way and therefore contravenes the axiom.  It is a fact that no function which falls faster than $\exp(-a \tau)$ at large $\tau$, where $a$ is some constant, can be expressed through Eq.\,(\ref{Widder}).\footnote{Correspondence with W.~N.~Polyzou was instrumental in developing this aspect of the discussion.}  Amongst other things, it follows that an excitation described by an effective mass, $m(\tau)$, which grows with Euclidean separation is necessarily absent from the spectrum of physical states because its Schwinger function violates the axiom of reflection positivity.  This is a novel perspective.

With the information provided above one can identify a confined excitation; and hence, in the context of QCD, these tools have been employed to analyse propagators obtained both as solutions of truncated systems of DSEs and through numerical simulations of lattice-regularised QCD.  Extant analyses, e.g., Ref.\,\cite{von Smekal:1997vx}, suggest strongly that the gluon $2$-point function in QCD is not reflection positive.  The same has been argued for the dressed-quark propagator in QCD, e.g., Ref.\,\cite{Bhagwat:2003vw}.  However, as noted above, the detailed structure of the dressed-quark-gluon vertex is unknown and it can influence the analytic properties of $S(p)$.  Hence a firm conclusion on the dressed-quark propagator has not yet been reached (see, e.g., Ref.\,\cite{Alkofer:2003jj}).

In relation to the studies just mentioned, it is noteworthy that any $2$-point Schwinger function with an inflexion point at $p^2>0$ must breach the axiom of reflection positivity.  This can be seen directly from Eq.\,(\ref{KLrep}), which constrains the first derivative of reflection positive functions to be negative definite and the second derivative to be positive definite.  Hence, it is often unnecessary to actually calculate the Fourier transform described in Eq.\,(\ref{Deltatau}).  A violation of reflection positivity can be determined by inspection of the pointwise behaviour of the momentum space Schwinger function.  

It should now be clear that we have a means of identifying whether or not a given Schwinger function can be associated with an expectation value in the Hilbert space of observables.  In principle, one could calculate all QCD's Schwinger functions and demonstrate confinement of colour therewith.  Plainly, however, that will be difficult in practice and nigh on impossible by brute force.  A proof of confinement at some higher organisational level is necessary, and sought by many.

There is a surviving hope that the study of gauge theories in less than four dimensions might be useful in this regard.  Analogous to quenched QCD, quenched QED in three dimensions (two spacial, one temporal -- QED$_3$) is confining because it has a nonzero string tension \cite{Gopfert:1981er}.  The effect of unquenching; viz., allowing light fermions to influence the theory's dynamics, remains an important but unresolved question.  It is conceivable that through its answer we might inform ourselves about aspects of the confinement problem in QCD.  The early history of QED$_3$ studies is reviewed in Ref.\,\cite{Roberts:1994dr} and their current status can be traced from Ref.\,\cite{bashirraya}.

In closing this section it is important to reiterate that confinement is a conjecture.  In order to establish confinement as more than a contemporary empirical fact, it will be necessary to map out the nonperturbative structure of QCD's $\beta$-function.\footnote{The requirement that renormalisation preserve gauge invariance; viz., that the renormalised QCD Lagrangian be invariant under Becchi-Rouet-Stora transformations, leads to the Slavnov-Taylor identities, which entail that all Schwinger functions which can be used to define the running coupling must yield the same result for that coupling.  It follows that QCD possesses a unique $\beta$-function.  For a detailed discussion, see, e.g., Ref.\,\cite{PT84}.}  In order to accomplish that, one will need to determine the nonperturbative evolution of the renormalisation constants.  That is tantamount to establishing rigorously that the theory exists.  This fact emphasises the magnitude of the problem.

Absent this, progress in hadron physics can still be made with well-constrained models that express key consequences of confinement.  The feedback that this enables between experiment and theory further illuminates the essence of the confinement conjecture in relation to light-quarks.  This can begin with the development and employment of a Poincar\'e covariant framework that ensures the colour-singlet $S$-matrix elements associated with physical processes do not possess production thresholds for light-quarks, in particular, and coloured elementary excitations in general.  The notions described above have been immensely valuable in this regard.  For example, they underly the DSE model introduced in Ref.\,\cite{Roberts:1994hh}, which has served as an archetype for numerous efficacious applications, as can be seen, e.g., from Refs.\,\cite{Roberts:2000aa,Alkofer:2000wg,Maris:2003vk,Roberts:2007jh}.

\section{Dynamical Chiral Symmetry Breaking}
Many statements of fact can be made in connection with this emergent phenomenon.  For example, DCSB explains the origin of constituent-quark masses and underlies the success of chiral effective field theory.  Understanding DCSB within QCD proceeds from the renormalised gap equation \cite{Maris:1997hd}:
\begin{equation}
S(p)^{-1} =  Z_2 \,(i\gamma\cdot p + m^{\rm bm}) + Z_1 \int^\Lambda_q\! g^2 D_{\mu\nu}(p-q) \frac{\lambda^a}{2}\gamma_\mu S(q) \Gamma^a_\nu(q,p) , \label{gendse}
\end{equation}
where $\int^\Lambda_q$ represents a Poincar\'e invariant regularisation of the integral, with $\Lambda$ the regularisation mass-scale, $D_{\mu\nu}$ is the renormalised dressed-gluon propagator, $\Gamma_\nu$ is the renormalised dressed-quark-gluon vertex, and $m^{\rm bm}$ is the quark's $\Lambda$-dependent bare current-mass.  The vertex and quark wave-function renormalisation constants, $Z_{1,2}(\zeta^2,\Lambda^2)$, depend on the gauge parameter.  

\begin{figure}[t]
\begin{center}

\includegraphics[clip,width=0.6\textwidth]{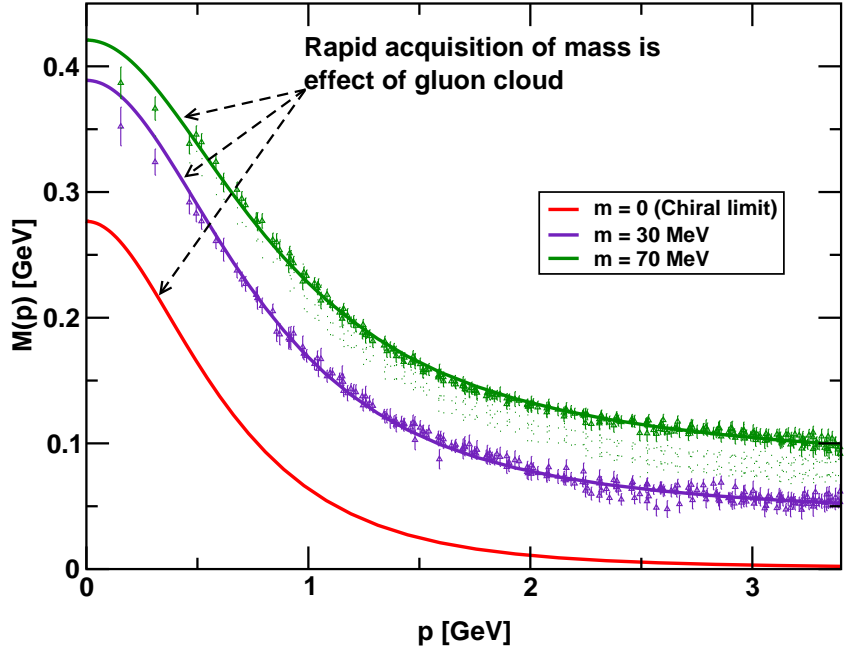}

\parbox{16.5cm}{\caption{\label{gluoncloud} Dressed-quark mass function, $M(p)$: solid curves -- DSE results, obtained as explained in Refs.\,\protect\cite{Bhagwat:2003vw,Bhagwat:2006tu}, ``data'' -- numerical simulations of unquenched lattice-QCD \protect\cite{Bowman:2005vx}.  In this figure, adapted from Ref.\,\protect\cite{Bhagwat:2007vx}, one observes the current-quark of perturbative QCD evolving into a constituent-quark as its momentum becomes smaller.  The constituent-quark mass arises from a cloud of low-momentum gluons attaching themselves to the current-quark.  This is dynamical chiral symmetry breaking: an essentially nonperturbative effect that generates a quark mass \emph{from nothing}; namely, it occurs even in the chiral limit.}}
\end{center}
\end{figure}

The form of the solution to Eq.\,(\ref{gendse}) is the same as that written in Eq.\,(\ref{SpAB}) \emph{except} that in QCD one must account for the renormalisation point dependence:
\begin{eqnarray} 
 S(p) & =&  -i \gamma\cdot p \,\sigma_V(p^2,\zeta^2) + \sigma_S(p^2,\zeta^2) = \frac{1}{i \gamma\cdot p \, A(p^2,\zeta^2) + B(p^2,\zeta^2)} =
\frac{Z(p^2,\zeta^2)}{i\gamma\cdot p + M(p^2)}\,.
%
\label{Sgeneral}
\end{eqnarray} 
It is important that the mass function, $M(p^2)=B(p^2,\zeta^2)/A(p^2,\zeta^2)$, illustrated in Fig.\,\ref{gluoncloud}, is independent of the renormalisation point, $\zeta$.  The dressed propagator is obtained from Eq.\,(\ref{gendse}) augmented by the renormalisation condition\footnote{Owing to asymptotic freedom it is natural to fix the dressed-propagator to be free-particle-like at some large spacelike momentum scale, $\zeta$.}
\begin{equation}\label{renormS} 
\left.S(p)^{-1}\right|_{p^2=\zeta^2} = i\gamma\cdot p +
m(\zeta^2)\,,
\end{equation}
where $m(\zeta^2)$ is the running mass: 
\begin{equation}
Z_2(\zeta^2,\Lambda^2) \, m^{\rm bm}(\Lambda) = Z_4(\zeta^2,\Lambda^2) \, m(\zeta^2)\,,
\end{equation}
with $Z_4$ the Lagrangian-mass renormalisation constant.  In QCD the chiral limit is strictly defined by \cite{Maris:1997hd}:
\begin{equation}
\label{limchiral}
Z_2(\zeta^2,\Lambda^2) \, m^{\rm bm}(\Lambda) \equiv 0 \,, \forall \Lambda^2 \gg \zeta^2 ,
\end{equation}
which states that the renormalisation-point-invariant cur\-rent-quark mass $\hat m = 0$.  

In perturbation theory it is impossible in the chiral limit to obtain $M(p^2)\neq 0$: the generation of mass \emph{from nothing} is an essentially nonperturbative phenomenon.  On the other hand, it is a longstanding prediction of nonperturbative DSE studies that DCSB will occur so long as the integrated infrared strength possessed by the gap equation's kernel exceeds some critical value \cite{Roberts:1994dr}.  There are strong indications that this condition is satisfied in QCD \cite{Bhagwat:2006tu,Bowman:2005vx} (see Fig.\,\ref{gluoncloud}).  It follows that the quark-parton of QCD acquires a momentum-dependent mass function, which at infrared momenta is $\sim 100$-times larger than the current-quark mass.  This effect owes primarily to a dense cloud of gluons that clothes a low-momentum quark \cite{Bhagwat:2007vx}.  It means that the Higgs mechanism is largely irrelevant to the bulk of normal matter in the universe.  Instead the single most important mass generating mechanism for light-quark hadrons is the strong interaction effect of DCSB; e.g., one can identify it as being responsible for 98\% of a proton's mass. 

The vacuum quark condensate is the order parameter most commonly cited in connection with DCSB.  For example, it occurs as a parameter at leading order in chiral perturbation theory.  It is a basic fact that only in the chiral limit is it possible to unambiguously define the gauge invariant vacuum quark condensate in terms of $S(p)$ \cite{Maris:1997hd,Langfeld:2003ye,Chang:2006bm}.  That such a definition is possible at all emphasises that gauge covariant quantities contain gauge invariant information.  Notwithstanding these truths, $M(p^2)$ is a more fundamental tracer for DCSB.  The condensate is only a small part of the information that the mass function contains, amongst other pieces are pointwise consequences of Goldstone's theorem \cite{Maris:1997hd}, definitions and an explanation of the constituent-quark mass \cite{Bhagwat:2007vx,Maris:1997tm,Bhagwat:2006xi}, and a related $\sigma$-term \cite{Holl:2005st}.\footnote{The current-quark mass dependence of both this $\sigma$-term and the massive-quark condensate defined unambiguously in Ref.\,\cite{Chang:2006bm} demonstrate that the essentially dynamical component of chiral symmetry breaking decreases with increasing current-quark mass.  This is an important observation that is basic, e.g., to the validity of heavy-quark effective theory.}  Harking back to Sect.\,\ref{Sect:Conf}, it is interesting to note that considered as a function of current-quark mass the dressed-quark propagator possesses a spectral representation \cite{Langfeld:2003ye}, confinement and DCSB notwithstanding.

That a nonzero quark mass function is generated in massless QCD entails that chiral symmetry is realised in the Nambu-Goldstone mode and the vacuum quark condensate is nonzero.  The magnitude of the mass function in the infrared and order parameters contingent thereupon underly the validity of chiral effective theory as a tool for correlating low-energy observables in QCD.  The question of whether $M(p^2)$ has an expansion in current-quark mass around its chiral-limit value bears upon the radius of convergence for that perturbation theory.  This question asks whether it is possible to write
\begin{equation}
M(p^2;\hat m) = M(p^2;\hat m=0) + \sum_{n=1}^\infty a_n \hat m^n
\end{equation}
on a measurable domain of current-quark mass.  It was found \cite{Chang:2006bm} that such an expansion exists for $\hat m < \hat m_{\rm rc}$ and is absolutely convergent.  The value of $\hat m_{\rm rc}$ can be reported as follows:\footnote{NB.\ Irrespective of the current-mass of the other constituent, a pseudoscalar meson containing one current-quark whose mass exceeds $\hat m_{\rm cr}$ is never within the domain of absolute convergence.} for a pion-like meson constituted from a quark, $f$, with mass $\hat m_{\rm rc}$ and an equal-mass but different flavor antiquark, $\bar g$, $m_{\bar g f}^{0^-} = 0.45\,$GeV.  Since physical observables, such as the leptonic decay constant, are expressed via $M(p^2)$, it follows that a chiral expansion is meaningful only for $(m_{\bar g f}^{0^-})^2 \lesssim 0.2\,$GeV$^2$.  Hence, it is only valid to employ chiral perturbation theory to fit and extrapolate results from numerical simulations of lattice-regularized QCD when the simulation parameters provide for $m_\pi^2 \lesssim 0.2\,$GeV$^2$.  Lattice results at larger pion masses are not within the domain of convergence of chiral perturbation theory. 

\section{Axial-vector Ward-Takahashi identity and mesons}
Dynamical chiral symmetry breaking is a fact in QCD and this amplifies the importance of the axial-vector Ward-Takahashi identity.  The existence of a sensible truncation of the DSEs \cite{Munczek:1994zz,Bender:1996bb} has enabled proof via that identity of a body of exact results for pseudoscalar mesons.  They relate even to radial excitations and/or hybrids \cite{Holl:2004fr,Holl:2005vu,McNeile:2006qy}, and heavy-light \cite{Ivanov:1997yg,Ivanov:1998ms} and heavy-heavy mesons \cite{Bhagwat:2006xi}.  The results have been illustrated using a renormalisation-group-improved rainbow-ladder truncation \cite{Maris:1997tm,Maris:1999nt}, which also provided a prediction of the electromagnetic pion form factor \cite{Maris:2000sk}.  In addition, algebraic parametrisations of the dressed-quark propagators and meson bound-state amplitudes obtained from such studies continue to be useful.  One example is a recent application to $B$-meson $\to$ light-meson transition form-factors \cite{Ivanov:2007cw}, the results of which can be useful in the analysis and correlation of the large body of data being accumulated at extant heavy-quark facilities, and thereby in probing the Standard Model.

\subsection{Neutral pseudoscalars and the \mbox{\boldmath $\eta^\prime$}} Implications for neutral pseudoscalar mesons have been elucidated \cite{Bhagwat:2007ha}.  In the general case the axial-vector Ward-Takahashi identity is written 
\begin{equation}
P_\mu \Gamma_{5\mu}^a(k;P) ={\cal S}^{-1}(k_+) i \gamma_5 {\cal F}^a 
+ i \gamma_5 {\cal F}^a {\cal S}^{-1}(k_-)
- 2 i {\cal M}^{ab}\Gamma_5^b(k;P)  - {\cal A}^a(k;P)\,,
\label{avwti}
\end{equation}
wherein: $\{{\cal F}^a | \, a=0,\ldots,N_f^2-1\}$ are the generators of $U(N_f)$; the dressed-quark propagator ${\cal S}=\,$diag$[S_u,S_d,S_s,S_c,S_b,\ldots]$ is matrix-valued; ${\cal M}(\zeta)$ is the matrix of renormalised (running) current-quark masses and
${\cal M}^{ab} = {\rm tr}_F \left[ \{ {\cal F}^a , {\cal M} \} {\cal F}^b \right],$
where the trace is over flavour indices.  The inhomogeneous axial-vector vertex in Eq.\,(\ref{avwti}) satisfies a Bethe-Salpeter equation; viz., 
\begin{equation}
\left[\Gamma^a_{5\mu}(k;P)\right]_{tu}
=  Z_2 \left[\gamma_5\gamma_\mu {\cal F}^a \right]_{tu}+ \int^\Lambda_q
[{\cal S}(q_+) \Gamma^a_{5\mu}(q;P) {\cal S}(q_-)]_{sr} K_{tu}^{rs}(q,k;P)\,,
\label{avbse}
\end{equation}
where: $P$ is the total momentum and $k$, $q$ are relative momenta, $k_\pm=k\pm P/2$; $r$,\ldots,\,$u$ represent colour, Dirac and flavour indices; and the quantity $K(q,k;P)$ is the Bethe-Salpeter kernel.  The pseudoscalar vertex $\Gamma_{5}^a$ satisfies an analogous equation driven by the inhomogeneity $Z_4 \gamma_5 {\cal F}^a$.  

The final term in Eq.\,(\ref{avwti}) expresses the non-Abelian axial anomaly.  It involves 
\begin{equation}
{\cal A}_U(k;P) = \!\!  \int\!\! d^4xd^4y\, e^{i(k_+\cdot x - k_- \cdot y)} N_f \left\langle  {\cal F}^0\,q(x)  \, {\cal Q}(0) \,   \bar q(y) 
\right\rangle, \label{AU}
\end{equation}
wherein the matrix element represents an operator expectation value in full QCD and
\begin{equation}
{\cal Q}(x) = i \frac{\alpha_s }{8 \pi} \, \epsilon_{\mu\nu\rho\sigma} F^a_{\mu\nu} F^a_{\rho\sigma}(x) 
= \partial_\mu K_\mu(x)
\end{equation}
is the topological charge density operator, with $F_{\mu\nu}^a$ the gluon field strength tensor.  It is fundamentally important that while ${\cal Q}(x)$ is gauge invariant, the associated Chern-Simons current, $K_\mu$, is not.  Thus in QCD no physical state can couple to $K_\mu$.  Hence, physical states cannot provide a resolution of the so-called $U_A(1)$-problem; i.e., they cannot play any role in ensuring that the $\eta^\prime$ is not a Goldstone mode.

It is plain and important that only ${\cal A}^{a=0}$ can be nonzero.  Moreover, in considering the $U_A(1)$-problem one need only focus on the case ${\cal A}^{0} \neq 0$ because if that is false, then following Ref.\,\cite{Maris:1997hd} it is clear that the $\eta^\prime$ is certainly a Goldstone mode.  ${\cal A}^0$ is a pseudoscalar vertex and can therefore be expressed
\begin{equation}
{\cal A}^0(k;P) = {\cal F}^0\gamma_5 \left[ i {\cal E}_{\cal A}(k;P) + \gamma\cdot P {\cal F}_{\cal A}(k;P)   +\gamma\cdot k k\cdot P {\cal G}_{\cal A}(k;P) + \sigma_{\mu\nu} k_\mu P_\nu {\cal H}_{\cal A}(k;P)\right].
\end{equation}
Equation\,(\ref{avwti}) can now be used to derive a collection of chiral-limit, pointwise Goldberger-Treiman relations, important amongst which is the identity
\begin{equation}
\label{ewti}
2 f_{\eta^\prime} E_{\eta^\prime}(k;0) = 2 B_{0}(k^2) - {\cal E}_{\cal A}(k;0)\,,
\end{equation}
where $B_0(k^2)$ is obtained in solving the chiral-limit gap equation.  

It is now plain that if 
\begin{equation}
\label{calEB}
{\cal E}_{\cal A}(k;0) = 2 B_{0}(k^2) \,,
\end{equation}
then $f_{\eta^\prime} E_{\eta^\prime}(k;0) \equiv 0$.  This being true, then the homogeneous Bethe-Salpeter equation for the $\eta^\prime$ does not possess a massless solution in the chiral limit.  The converse is also true; namely, the absence of such a solution requires Eq.\,(\ref{calEB}).  Hence,  Eq.\,(\ref{calEB}) is a necessary and sufficient condition for the absence of a massless $\eta^\prime$ bound-state.  It is the chiral limit that is being discussed, in which case $B_{0}(k^2) \neq 0$ if, and only if, chiral symmetry is dynamically broken.   Hence, the absence of a massless $\eta^\prime$ bound-state is only assured through the existence of an intimate connection between DCSB and an expectation value of the topological charge density.  A  relationship between the mechanism underlying DCSB and the absence of a ninth Goldstone boson was also discussed in Ref.\,\cite{Dorokhov:2003kf}.

Reference\,\cite{Bhagwat:2007ha} derives a range of corollaries, amongst which are mass formulae for neutral pseudoscalar mesons.  These led, e.g., to a prediction of the manner by which the $\eta^\prime$ is split from the octet pseudoscalars by an amount that depends on QCD's topological susceptibility.  That is most easily illustrated by considering the $U(N_f)$ limit, in which all current-quark masses assume the single value $m(\zeta)$.  In this case one finds 
\begin{equation}
\label{etapchiral}
m_{\eta^\prime}^2 f_{\eta^\prime}^0 = n_{\eta^\prime} + 2 m(\zeta)\rho_{\eta^\prime}^0(\zeta) \,,
\end{equation}
where $m_{\eta^\prime}$ is the meson's mass,
\begin{eqnarray} 
\label{fpiacpres} f_{\eta^\prime}^0 \,  P_\mu = Z_2\,{\rm tr} \int^\Lambda_q 
{\cal F}^0 \gamma_5\gamma_\mu\, \chi_{\eta^\prime}(q;P) \,, 
&\;&
i  \rho_{\eta^\prime}^0\!(\zeta)  = Z_4\,{\rm tr} 
\int^\Lambda_q {\cal F}^0 \gamma_5 \, \chi_{\eta^\prime}(q;P)\,,
\end{eqnarray} 
with $\chi_{\pi_i}(k;P) = {\cal S}(k_+) \Gamma_{\pi_i}(k;P) {\cal S}(k_-)$, and the residue of the $\eta^\prime$ bound-state pole in  ${\cal A}^0$ of Eq.\,(\ref{AU}) is
\begin{equation}
n_{\eta^\prime} = \mbox{\footnotesize $\displaystyle \sqrt{\frac{N_f}{2}}$} \, \nu_{\eta^\prime} \,, \; \nu_{\pi_i}= \langle 0 | {\cal Q} | \eta^\prime\rangle \,.
\end{equation}
Plainly, the $\eta^\prime$ is split from the Goldstone modes so long as $n_{\eta^\prime} \neq 0$.\footnote{A nonzero value of the topological susceptibility can be achieved through the coupling of a massless axial-vector gauge-field ghost to the Chern-Simons current, which does not appear in the particle spectrum of QCD because the current is not gauge invariant.  This is a variant of the Kogut-Susskind mechanism \protect\cite{Kogut:1974kt}.}  

It is argued \cite{Witten:1979vv,Veneziano:1979ec} that in QCD 
\begin{equation}
n_{\eta^\prime} \sim \frac{1}{\sqrt{N_c}}\,,
\end{equation}
and it can be seen to follow from the gap equation, the homogeneous BSE and Eqs.\,(\ref{fpiacpres}) that 
\begin{equation}
f_{\eta^\prime}^0 \sim \sqrt{N_c} \sim \rho_{\eta^\prime}^0(\zeta)\,.
\end{equation}
Consider now Eq.\,(\ref{etapchiral}) in the form
\begin{equation}
m_{\eta^\prime}^2 =  \frac{n_{\eta^\prime}}{f_{\eta^\prime}^0} + 2 m(\zeta) \frac{\rho_{\eta^\prime}^0(\zeta)}{f_{\eta^\prime}^0} \,.
\end{equation}
The first term is zero in the limit $N_c\to \infty$ while the second remains finite.  Subsequently taking the chiral limit, $m_{\eta^\prime}$ vanishes in the manner characteristic of all Goldstone modes.\footnote{NB.\ One must take the limit $N_c\to \infty$ before the chiral limit because the procedures do not commute \cite{Narayanan:2004cp}.} These results are realised in the effective Lagrangian of Ref.\,\cite{Di Vecchia:1979bf} in a fashion that is consistent with all the constraints of the anomalous Ward identity.  This is \emph{not} true of the so-called 't\,Hooft determinant \protect\cite{Crewther:1977ce,Crewther:1978zz,Christos:1984tu}.

Reference\,\cite{Bhagwat:2007ha}  also presents an \textit{Ansatz} for the Bethe-Salpeter kernel that enables illustration of the implications of the mass formulae, in particular mixing between neutral pseudoscalars.\footnote{Interplay between the non-Abelian anomaly and flavor symmetry breaking is also explored in Ref.\,\protect\cite{Kekez:2005ie}.}   Despite its simplicity, the model is elucidative and phenomenologically efficacious; e.g., it predicts $\eta$--$\eta^\prime$ mixing angles of $\sim - 15^\circ$, $\pi^0$--$\eta$ angles of $\sim 1^\circ$, and a strong neutron-proton mass difference of $0.75\,(\hat m_d - \hat m_u)$.

\subsection{Meson spectroscopy}
The use of DSEs to study meson phenomena is empowered by the existence noted above of a systematic, nonperturbative and symmetry-preserving truncation scheme \cite{Munczek:1994zz,Bender:1996bb}.  It means that exact results, such as those outlined heretofore, can be illustrated and predictions for experiment made with readily quantifiable errors.  The renormalisation-group-improved rainbow-ladder truncation of the gap and Bethe-Salpeter equations introduced in Refs.\,\cite{Maris:1997tm,Maris:1999nt} has been widely employed in this endeavour.  To exemplify that, in Table\,\ref{masses} calculated results are reported for the masses of the lightest $J=0,1$ states \cite{Cloet:2007pi}.  It is true in general that the truncation is accurate for the $0^{-+}$ and $1^{--}$ light-quark meson ground states.  In these channels it can be seen algebraically that contributions beyond rainbow-ladder largely cancel between themselves owing to Eq.\,(\ref{avwti}) \cite{Bhagwat:2004hn,Matevosyan:2006bk,Bender:1996bb,Bender:2002as}.  The remaining columns in the table deserve special attention because they show clearly the path toward improvement.

\begin{table}[t]
\begin{center}
\begin{minipage}[t]{16.5 cm}
\caption{\label{masses} Masses (GeV) of the lightest $J=0,1$ states produced by the rainbow-ladder DSE truncation of Refs.\,\protect\cite{Maris:1997tm,Maris:1999nt} with the parameter values: $\omega = 0.4\,$GeV, $\omega D = (0.72\,$GeV$)^3$; and current-quark masses $m_{u,d}(1\,{\rm GeV}) = 5.45\,$MeV, $m_{s}(1\,{\rm GeV}) = 125\,$MeV.  The rainbow-ladder kernel gives ideal flavour mixing for all states, and hence the pure $\bar s s$ $0^{-+}$ state is an artefact which does not correspond to a physical state.  A simple kernel that implements flavour mixing is described in Ref.\,\protect\cite{Bhagwat:2007ha}. 
The text explains in detail the results in this table, which is adapted from Ref.\,\protect\cite{Cloet:2007pi}.\smallskip
}
\end{minipage}
\begin{tabular}{l||l|l||l||l|l||l|l|l} \hline \hline
\rule{0ex}{3ex} $J^{PC}$ & $0^{-+}$ & $1^{--}$ & $0^{++}$ & $1^{+-}$ & $1^{++}$ & $0^{--}$ & $1^{-+}$ & $0^{+-}$ \\
$\bar u u$ & 139 & ~740 & ~670 & ~830  & ~900 & ~860 & 1000 & 1040 \\
$\bar s s$ & 695 & 1065 & 1080 & 1165 & 1240 & 1170 & 1310 & 1385
\\
\hline \hline
\end{tabular} 
\end{center}
\end{table}

Terms beyond the rainbow-ladder truncation are known to add constructively in the $0^{++}$ channel \cite{Roberts:1996jx}.  Hence the leading order truncation is \emph{a priori} not expected to provide a good approximation.  Further understanding is provided by an exploration of the contribution from two-pion intermediate states to the mass and width of this lowest-mass scalar.  A rudimentary analysis shows that a realistic description is attainable therewith \cite{Holl:2005st}; viz., it gives a pole position $\surd s_\sigma = (0.578 - i \, 0.311)\,$GeV.  This is not the end of the scalar story but it is a sensible path to follow, in particular because a QCD-level mechanism is precisely specified.  Something analogous should be considered in connection with the $\bar s s$ scalar state listed in Table~\ref{masses}.  It is notable that in rainbow-ladder truncation it has been shown that at least up to the c-quark mass the ordering of meson masses is $0^{-+}(1S) < 0^{++}(1S) < 0^{-+}(2S) < 0^{++}(2S)$, where $n=1,2$ denotes the ground state and first radial excitation, respectively \protect\cite{Krassnigg:2006ps}.

Compared with experiment, the masses of the axial-vector mesons $1^{+\pm}$ are poorly described by the rainbow-ladder truncation: $\sim 400\,$MeV of repulsion is missing from the kernel.  A cruder model does better \cite{Burden:1996nh,Bloch:1999vka}.  The latter studies and a more recent analysis \cite{Watson:2004kd} indicate that at least part of the defect owes to the absence of spin-flip contributions at leading-order.  Such contributions appear at all higher orders and are enhanced by the strongly dressed quark mass function. This is one of the ways that the meson spectrum can be used to probe the long-range part of the light-quark interaction and thereby to chart the nonperturbative behavior of QCD's $\beta$-function. 

The last three columns describe systems with so-called exotic quantum numbers.  Of course, these states are exotic only in the context of the naive constituent quark model.  In QCD they correspond simply to interpolating fields with some gluon content and are easily accessible via the BSE \cite{Burden:2002ps}.  Nonetheless, while the rainbow-ladder truncation binds in these channels, the shortcomings encountered in the $1^{+}$ channels are also evident here, for many of the same reasons.  Reliable predictions for the masses of such states will only be obtained once improved kernels are developed.  At the very least, one must have dependable predictions for axial-vector masses before drawing any conclusions about the so-called exotics.  Since all these states lie within a domain on which DCSB is very relevant, a framework with a veracious description of that emergent phenomenon is essential. 

One might pose the question of whether, in the context of bound-state studies in which model assumptions are made regarding the nature of the long-range interaction between light quarks, anything is gained by working solely with Schwinger functions.  This means, in part, constraining oneself to work only with information obtained from the DSEs at spacelike momenta.\footnote{Lattice-regularised QCD provides a background to this question.  That approach is grounded on the Euclidean space functional integral.  Schwinger functions are all that it can directly provide.  Hence, lattice-regularised QCD can only be useful if methods are found so that the question can be answered in the affirmative.}  According to a recent study \cite{Bhagwat:2007rj} the answer is no.  It analysed the capacity of Schwinger functions to yield information about bound states, and established that for the ground state in a given channel the mass and residue are accessible via rudimentary methods.  However, simple methods cannot provide dependable information about more massive states in a given channel.  Indeed, there is no easy way to extract such information.  An approach based on a correlator matrix can be successful but only if the operators are carefully constructed so as to have large overlap with states of interest in a given channel, and statistical and systematic errors can be made small; viz., $\sim 1$\%.  While it is possible in principle to satisfy these constraints, doing so is labor intensive and time consuming.  That is only justified in the absence of model-dependence.

\section{Nucleons}
Partly because attention to this sector has only recently increased, the current level of expertise in studying baryons is roughly the same as it was with mesons more than ten years ago; viz., model building and phenomenology.  We are a little ahead of that game, however, because much has been learnt in meson applications; e.g., as indicated above, a veracious understanding of the structure of dressed-quarks and -gluons has been acquired.   

\subsection{Faddeev equation}
In quantum field theory a nucleon appears as a pole in a six-point quark Green function.  The pole's residue is proportional to the nucleon's Faddeev amplitude, which is obtained from a Poincar\'e covariant Faddeev equation that adds-up all possible quantum field theoretical exchanges and interactions that can take place between three dressed-quarks.  This is important because modern, high-luminosity experimental facilities employ large momentum transfer reactions; viz., $Q^2 > M_N^2$ where $M_N$ is the nucleon's mass, and hence a veracious understanding of contemporary data requires a Poincar\'e covariant description of the nucleon.  

A tractable truncation of the Faddeev equation is based \cite{Cahill:1988dx} on the observation that an interaction which describes mesons also generates diquark correlations in the colour-$\bar 3$ channel \cite{Cahill:1987qr}.  The dominant correlations for ground state octet and decuplet baryons are $0^+$ and $1^+$ diquarks because, e.g.: the associated mass-scales are smaller than the baryons' masses \cite{Burden:1996nh,Maris:2002yu}, namely (in GeV) -- 
\begin{equation}
\label{diquarkmass}
m_{[ud]_{0^+}} = 0.7 - 0.8
 \,,\; 
m_{(uu)_{1^+}}=m_{(ud)_{1^+}}=m_{(dd)_{1^+}}=0.9 - 1.0\,;
\end{equation}
and the electromagnetic size of these correlations, while larger than the pion, is less than that of the proton \cite{Maris:2004bp} -- 
\begin{equation}
r_{[ud]_{0^+}} \approx 0.7\,{\rm fm}\,,\; r_{(ud)_{1^+}} \sim 0.8\,{\rm fm}\,.
\end{equation}
The last result is an estimate based on the $\rho$-meson/$\pi$-meson radius-ratio \cite{Maris:2000sk,Bhagwat:2006pu}.

The Faddeev equation's kernel is completed by specifying that the quarks are dressed, with two of the three dressed-quarks correlated always as a colour-$\bar 3$ diquark.  Binding is then effected by the iterated exchange of roles between the bystander and diquark-participant quarks.  A Ward-Takahashi-identity-pre\-ser\-ving electromagnetic current for the baryon thus constituted is subsequently derived~\cite{Oettel:1999gc}.  It depends on the electromagnetic properties of the axial-vector diquark correlation.

\subsection{Nucleon form factors}
A study of the nucleon's mass and the effect on this of a pseudoscalar meson cloud are detailed in \cite{Hecht:2002ej}.  Lessons learnt were employed in a series of studies of nucleon properties, including form factors \cite{Alkofer:2004yf,Holl:2005zi,Bhagwat:2006py}.  The calculated ratio $\mu_p G_E^p(Q^2)/G_M^p(Q^2)$ passes through zero at $Q^2\approx 6.5\,$GeV$^2$ \cite{Holl:2005zi}.  For the neutron, in the neighbourhood of $Q^2=0$, $\mu_n\, G_E^n(Q^2)/G_M^n(Q^2) = - \frac{r_n^2}{6}\, Q^2$, where $r_n$ is the neutron's electric radius \cite{Bhagwat:2006py}.  The evolution of $\mu_p G_E^p(Q^2)/G_M^p(Q^2)$ and $\mu_n G_E^n(Q^2)/G_M^n(Q^2)$ on $Q^2\gtrsim 2\,$GeV$^2$ are both primarily determined by the quark-core of the nucleon.  While the proton ratio decreases uniformly on this domain \cite{Alkofer:2004yf,Holl:2005zi}, the neutron ratio increases steadily until $Q^2\simeq 8\,$GeV$^2$ \cite{Bhagwat:2006py}.  

\begin{figure}[t]
\begin{center}

\includegraphics[clip,width=0.6\textwidth]{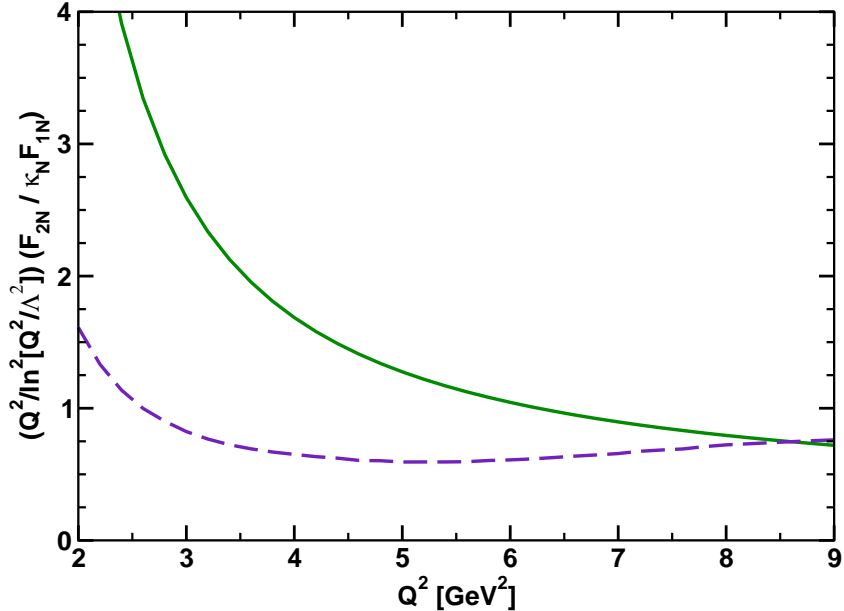}

\parbox{16.5cm}{\caption{\label{nF2F1} Weighted nucleon Pauli/Dirac form factor ratio, calculated using the framework of Ref.\,\protect\cite{Alkofer:2004yf} and presented with $\Lambda=0.94$GeV: solid curve -- neutron; dashed curve -- proton.  (Figure adapted from Ref.\,\protect\cite{Bhagwat:2007vx}.)
}}
\end{center}
\end{figure}

A form factor ratio motivated by ideas from perturbative QCD \cite{Belitsky:2002kj} is depicted in Fig.\,\ref{nF2F1}.  The parameter $\Lambda$ is interpreted as a mass-scale that defines the upper-bound on the domain of so-called soft momenta in the perturbative analysis.  A plausible value for such a quantity is $\Lambda \sim M_N$ \cite{Alkofer:2004yf}.\footnote{NB.\ A value of $\Lambda \sim 0.3\,$GeV corresponds to a length-scale $r_\Lambda \sim 1\,$fm.  It is not credible that perturbative QCD is applicable at ranges greater than the proton's radius.  A reasonable value is $\Lambda \sim 1\,$GeV, leading to $r_{1\,{\rm GeV}}\sim 0.2\,$fm.}  It is interesting that the model of Ref.\,\cite{Alkofer:2004yf} yields neutron and proton ratios which cross at $Q^2\simeq 8.5\,$GeV$^2$.  The model's prediction for truly asymptotic momenta is currently being explored.

It is plausible that Nature's fundamental ``constants'' might actually exhibit spatial and temporal variation \cite{Uzan:2002vq}.  This motivates an exploration of the current-quark-mass-dependence of the nucleon magnetic moments, which complements work on hadron $\sigma$-terms \cite{Holl:2005st,Flambaum:2005kc,Flambaum:2007mj}.  Preliminary results for the quark-core contribution to this variation can be expressed through the following ratios evaluated at the physical current-quark mass \cite{Bhagwat:2007vx}:
\begin{eqnarray}
-\frac{\delta \mu_p}{\mu_p} = 0.016 \, \frac{\delta m}{m}\,,&& 
-\frac{\delta \mu_n}{\mu_n} = 0.0042\, \frac{\delta m}{m} \,.
\end{eqnarray}
It is likely that pseudoscalar meson contributions will increase these values by a factor of $\gtrsim 10$ \cite{Flambaum:2004tm}.

The framework can naturally be applied to calculate weak and strong form factors of the nucleon.  Preliminary studies of this type are reported in Refs.\,\cite{jacquesmyriad,oettel2,Holl:2006zw}.  Such form factors are sensitive to different aspects of quark-nuclear physics and should prove useful, e.g., in constraining coupled-channel models for medium-energy production reactions on the nucleon.  This is important to the search for the so-called missing nucleon resonances\footnote{This refers to the fact that numerous excited states of the nucleon ($N^*$), which are predicted by the $SU(6)\otimes O(3)$ constituent quark model ($CQM_{6\times 3}$), are not seen in the baryon spectrum that is obtained from amplitude analyses of $\pi N$ elastic scattering.  At this time more than half of the low-lying states predicted by $CQM_{6\times 3}$ are missing \protect\cite{Lee:2006xu}.}  and the related problem of identifying exotic baryons.  

\subsection{Quark orbital angular momentum}
Of significant interest is the distribution of an hadron's \emph{spin} over the quark constituents and their angular momentum.  In a Poincar\'e covariant approach that can be calculated in any frame.  The rest frame is physically most natural.  The pion was considered in Ref.\,\cite{Bhagwat:2006xi} and although $J=0$ the dressed-quarks carry significant orbital angular momentum.  A mixing angle of $\sim 30^\circ$ can be attributed in the rest frame to the $L=1$ components of the pion's Poincar\'e covariant Bethe-Salpeter wave function.  

The answer is more complicated for the spin-$\frac{1}{2}$ nucleon.  In the truncation described above, a nucleon's Faddeev wave-function is expressed through eight scalar functions: no more are needed, no number fewer is complete.  Two are associated with the $0^+$ diquark correlation: ${\cal S}_{1,2}$, and six with the $1^+$ correlation: ${\cal A}_{1,\ldots,6}$.  In the rest frame in this basis one can derive the following ``good'' angular momentum and spin assignments, which add vectorially to give a $J=\frac{1}{2}$ nucleon:\footnote{Equation~(3.35) or Ref.\,\protect\cite{Oettel:1998bk} contradicts Fig.\,6 of that reference.  Equation~(\protect\ref{LS}) herein describes the correct assignments.}
\begin{eqnarray}
&&
\begin{array}{c|c|c|c}
L=0\,, S=\frac{1}{2} & L=1\,,S=\frac{1}{2} &  L=1\,,S=\frac{3}{2} & L=2 \,, S=\frac{3}{2}\\[1ex]\hline
{\cal S}_1\,,{\cal A}_{2}\,, {\cal B}_1 & {\cal S}_2\,, {\cal A}_1\,, {\cal B}_2\,, & {\cal C}_2& {\cal C}_1
\end{array}\;, \\
&&
\begin{array}{cclcclcclccl}
 B_1 & =& \frac{1}{3} A_3 + \frac{2}{3} A_5\,, &
 B_2 & =& \frac{1}{3} A_4 + \frac{2}{3} A_6\,, &
 C_1 & =& A_3 - A_5\,, &
 C_2 & =& A_4 - A_6 \,.
\end{array} \label{LS}
\end{eqnarray}
These assignments are straightforward to demonstrate and understand; e.g, in the rest frame of a relativistic constituent quark model the ${\cal S}_{1,2}$ terms correspond, respectively, to the upper and lower components of the nucleon's spinor.

To exhibit the importance of the various $L$-$S$ correlations within the nucleon's Faddeev wave-function, Ref.\,\cite{Cloet:2007pi} reported the breakdown of contributions to the nucleon's canonical normalisation:\footnote{The entry in location ${\cal S}_1 \otimes {\cal S}_1$ indicates the integrated contribution associated with ${\cal S}_1^2$.  The entries are reweighted such that the sum of the squares of the entries equals one.  Positions without an entry are zero to two decimal places.}
\begin{equation}
\begin{array}{l|rrr|rrr|r|r}
& {\cal S}_1  &  {\cal A}_2  &  {\cal B}_1 &   {\cal S}_2  &  {\cal A}_1& 
  {\cal B}_2 &   {\cal C}_2 &   {\cal C}_1 \\\hline
 {\cal S}_1 & 0.62 & -0.01 & 0.07 & 0.25 &  &  &  & -0.02 \\
{\cal A}_2& -0.01 &  & -0.06 &  & 0.05 & 0.04 & 0.02 & -0.16 \\
{\cal B}_1 & 0.07 & -0.06 & -0.01 &  & 0.01 & 0.13 & -0.01 &  \\\hline
{\cal S}_2& 0.25 &  &  & 0.06 &  & &  &  \\
{\cal A}_1&  & 0.05 & 0.01 &  &  & -0.07 & -0.07 & 0.02 \\
{\cal B}_2 &  & 0.04 & 0.13 &  & -0.07 & -0.10 & -0.02 & 0.13 \\\hline
{\cal C}_2 &  & 0.02 & -0.01 &  & -0.07 & -0.02 & -0.11 & 0.37 \\\hline
{\cal C}_1 & -0.02 & -0.16 &  &  & 0.02 & 0.13 & 0.37 & -0.15
\end{array}\label{LSresult}
\end{equation} 
To illustrate how Eq.\,(\ref{LSresult}) should be understood, note that the largest single entry is associated with ${\cal S}_1 \otimes {\cal S}_1$, which represents the quark outside the scalar diquark correlation carrying all the nucleon's spin.  That is the $u$-quark in the proton.  However, it is noteworthy that a contribution of similar magnitude is associated with the axial-vector diquark correlations, which express mixing between $p$- and $d$-waves; viz., ${\cal C}_1 \otimes {\cal C}_2 + {\cal C}_2 \otimes {\cal C}_1$.  With ${\cal C}_2$ all quark spins are aligned with that of the nucleon and the unit of angular momentum is opposed: $(q\uparrow)(q\uparrow)(q\uparrow)(P\downarrow)$, while with ${\cal C}_1$ all quark spins are opposed and the two units of angular momentum are aligned: $(q\downarrow)(q\downarrow)(q\downarrow)(D\uparrow)$.  This contribution is more important than those associated with ${\cal S}_2$; namely, scalar diquark terms with the bystander quark's spin antiparallel.  Finally, for the present, in this context one single number is perhaps most telling: the contribution to the normalisation from $(L=0) \otimes (L=0)$ terms is only 37\% of the total.

\section{Coda}
The basic problem of hadron physics is to solve QCD.  In order to achieve this goal a joint effort from experiment and theory is required.  Our community now has a range of major facilities that are accumulating data, of unprecedented accuracy and precision, which pose important challenges for theory.  It is the feedback between experiment and theory that leads most rapidly to progress in understanding.  The opportunities for hadron physics will grow because upgraded and new facilities will appear on a five-to-ten-year time-scale.  

This volume provides an excellent illustration of the splendid experimental accomplishments of recent times, the achievements of theory in explaining and interpreting the observations, and some of the urgent open problems.  Two of the things that stand out are the diversity of theoretical approaches and the interplay between them that is necessary to draw valid conclusions from contemporary data.

One need not think too hard in order to list a number of the key open problems.  It is essential to understand the origin and the nature of confinement.  Asymptotic coloured states have not hitherto been observed, but is it a cardinal fact that they cannot?  As described herein, confinement is related to the analytic structure of elementary n-point functions in QCD and especially to their properties at infrared momenta.  Is it in the essence of non-Abelian gauge theories that they possess an Hilbert space of physical states that is reconstructed from only a small subset of all the theory's Schwinger functions?   In progressing toward an answer to the question of confinement it will be necessary to map out the infrared behaviour of QCD's $\beta$-function.  The upgraded and future facilities will provide data which will guide that process.   However, to make full use of that data, it will be necessary to have reliable Poincar\'e covariant theoretical tools which enable the study of hadrons in the mass range $1$-$2\,$GeV.  On this domain confinement and dynamical chiral symmetry breaking are both germane.

It is known that dynamical chiral symmetry breaking (DCSB); namely, the generation of mass \emph{from nothing}, does take place in QCD.  However, the origin of the interaction strength at infrared momenta that guarantees DCSB through the gap equation is unknown.  This ties confinement to DCSB.  Further to this connection, at nonzero temperature and chemical potential are the deconfinement and chiral symmetry restoration transitions simultaneous?  If so, why, under what conditions, and also, perhaps, in which class of theories?  

It is important to understand the relationship between parton properties on the light-front and the rest frame structure of hadrons.  This is a problem because, e.g., DCSB, an established keystone of low-energy QCD, has not been realised in the light-front formulation.  The obstacle is the constraint $k^+:=k^0+k^3>0$ for massive quanta on the light front \cite{Brodsky:1991ir}.  It is therefore impossible to make zero momentum Fock states that contain particles and hence the vacuum is trivial.  Only the zero modes of light-front quantisation can dress the ground state but little progress has been made with understanding just how that might occur.  Furthermore, parton distribution functions must be calculated in order to comprehend their content.  Parametrisation is insufficient.  It would be very interesting to know, e.g., how, if at all, the distribution functions of a Goldstone mode differ from those of other hadrons.  Calculation \cite{Hecht:2000xa} and experiment  \cite{Wijesooriya:2005ir} together are needed to address that issue.   

A last question in this brief survey, but certainly not the least, how do the properties of hadrons change within nuclei, and within very dense systems?  Absent an answer to this, one is not addressing the subject of nuclear physics.  The experimental archetype for this problem is the EMC effect.  Experiments have shown that in-nucleus parton distribution functions cannot simply be obtained by adding
together the distributions within the constituent nucleons.  A satisfactory understanding of this effect is still lacking, almost twenty-five years after its discovery.  Within this theme one also asks for information on: the equation of state for dense matter; the properties of the hadron to quark-gluon phase transition; and the evolution between and nature of the states of matter within compact astrophysical objects.  Plainly, in addressing this question one is embarking on a far reaching quest.  

There is ample evidence of the constructive feedback between theoretical approaches.  For example, Dyson-Schwinger equation studies and effective field theory inform calculations within lattice-regularised QCD, and this interplay will become increasingly powerful as the parameters of lattice-QCD simulations come closer to the physical domain.  Notwithstanding the investment in and progress that has been made with lattice-regularised QCD, for example, there is a pressing present and continuing need for model building and application in hadron physics.  No single so-called \emph{ab-initio} approach is currently applicable to all the phenomena attributed to QCD at all the length-scales that can be explored.  Therefore, understanding and synthesising the wealth of extant and forthcoming data requires the continued development of models.  Such tools answer an immediate need and provide a ready means to develop intuition about a complex system.  Naturally, a reasonable model should obey all relevant constraints that \emph{ab-initio} approaches and symmetries apply.  That being so, it can then provide a rational and flexible guide in connection with our rapidly changing experimental environment.

Continued investment in each of the nonperturbative tools that might provide insight into QCD is also necessary.  Perturbative QCD is understood but its application begs a key question: at just which scale does its employment become valid?  The answer can only be attained nonperturbatively.  A number of experiments are described as exhibiting precocious scaling; i.e., they are associated with cross-sections that appear to evolve with momentum according to perturbative QCD expectations.  However, only a comprehensive nonperturbative analysis can explain whether that is the result of an accidental cancellation over the -- often small -- momentum domain accessible to the experiment or a clear signal of perturbative QCD.  Partial answers to this and related questions are possible through internally consistent model calculations that can unify many observables, e.g., Ref.\cite{Maris:1998hc}.  But a complete answer will require truly nonperturbative calculations of the scale-dependence of observables within a framework that is able to exhibit the perturbative QCD limit, e.g., Ref.\,\cite{Bhagwat:2006xi}.

This contribution has outlined the utilisation of DSEs in studying some of the many facets of the basic problem of hadron physics.  The DSEs provide a natural framework within which light-quark confinement can be defined, and the origin and consequences of DCSB explored.  To build an understanding of confinement, it is essential to work toward an accurate map of the interaction between light-quarks.  Among the rewards are a precise connection between confinement and DCSB, an accounting of the distribution of mass within hadrons, and a realistic picture of hybrids and exotics.  It cannot be too strongly emphasised that DCSB is remarkably effective at generating mass.  For light-quarks it is far more important than the Higgs mechanism.  It is understood via QCD's gap equation, which delivers a quark mass function with a momentum-dependence that connects the perturbative domain with the nonperturbative, con\-sti\-tuent-quark domain.  The existence of a sensible truncation scheme enables the proof of exact results using the DSEs.  The scheme is also tractable, and hence the results can be illustrated and predictions made for observables.  The consequent opportunities for rapid feedback between experiment and theory provide openings through which an intuitive understanding of numerous nonperturbative strong interaction phenomena can be reached via the DSEs.  

\section*{Acknowledgments}
I acknowledge valuable conversations with B.~El-Bennich, I.~C.~Clo\"et, T.~Kl\"ahn, A.~Krassnigg and R.~D.~Young.  
I am grateful to the organisers, and to the staff at the \emph{Ettore Majorana Foundation and Centre for Scientific Culture}, for their hospitality and support during the meeting.
This work was supported by the Department of Energy, Office of Nuclear Physics, contract no.\ DE-AC02-06CH11357; and benefited from the facilities of the ANL Computing Resource Center.

\end{document}